# Robust and effective *ab initio* molecular dynamics simulations on the GPU cloud infrastructure using the Schrödinger Materials Science Suite

Alexandr Fonari[a,*], Garvit Agarwal[a], Subodh C. Tiwari[a], Casey N. Brock[a], Jacob Gavartin[b], Mathew D. Halls[c]

[a] Schrödinger Inc., New York, NY 10036, United States
[b] Schrödinger Technologies Ltd, United Kingdom
[c] Schrödinger Inc., San Diego, CA 92121, United States

[*] Corresponding author. E-mail: alexandr.fonari[at]schrodinger.com. Address: 1540 Broadway, Floor 24, New York, NY 10036, US

## Abstract

*Ab initio* Born-Oppenheimer molecular dynamics (AIMD) is a valuable method for simulating physico-chemical processes of complex systems, including reactive systems, and for training machine learning models and force fields. Speed and stability issues on traditional hardware preclude routine AIMD simulations for larger systems and longer timescales. We postulate that any practically useful AIMD simulation must generate a trajectory of a minimum 1000 MD steps a day on a moderate cloud resource. In this work, we implement a computing workflow that enables routine calculations at this throughput and demonstrate results for several non-trivial atomistic dynamical systems. In particular, we have employed the GPU implementation of the Quantum ESPRESSO code which we will show increases AIMD productivity compared to the CPU version. In order to take advantage of transient servers (which are more cost and energy effective compared to the stable servers), we have implemented automatic restart/continuation of the AIMD runs within the Schrödinger Materials Science Suite. Finally, to reduce simulation size and thus reduce compute time when modeling surfaces, we have implemented a wall potential constraint. Our benchmarks using several reactive systems (lithium anode surface/solvent interface, hydrogen diffusion in an iron grain boundary) show a significant speed up when running on a GPU-enabled transient server using our updated implementation.

# 1. Introduction

Advancements of computational methods and computer power have led to rapid expansion of electronic structure modeling in materials science. Total energy methods are widely used for materials discovery, characterisation and screening problems, and generating a wealth of information complimentary to experiments and captured in publicly available databases such as Materials Project [1] and Materials Cloud [2]. While the vast utility of such databases is just beginning to make an impact in materials science, it is clear that *ab initio* molecular dynamics (AIMD) methods will have an increasingly larger role in atomistic modeling of complex multicomponent systems. There are several reasons for this:

1. Static relaxation methods are limited in probing the local potential energy landscape and chemical processes;
2. Classical molecular dynamics, especially for reactive species, requires accurate descriptions of interatomic forces that may be generated using AIMD;
3. Training machine learning models and force fields require accurate and reasonably long trajectories.

Cloud computing becomes readily available for researchers worldwide [3]. The cloud itself is composed of servers (nodes) that run jobs (calculations). Besides stable servers that are continuously available, transient (also known as preemptive or preemptible) servers have recently been introduced [4]. These servers are available only temporarily and can shut down at any time. Transient servers are deemed to be more energy efficient due to several reasons: (1) currently idling stable servers' resources can be repurposed for transient servers (and repurposed back at any time); (2) cloud infrastructure that relies on alternative power sources (solar/wind) that generate power intermittently is more suitable for transient servers [4]. Currently, several major cloud providers offer transient servers for the general public, including Amazon [5], Google [6], IBM [7], and Microsoft [8].

The Schrödinger Materials Science Suite [9] provides a comprehensive integration with Quantum ESPRESSO (QE) [10]. This includes a graphical user interface for calculation setup, job management and results analysis, as well as automated workflows for a variety of materials properties and common calculations. Precompiled QE binaries are provided to ease the user's installation. In addition to the already available CPU binaries, we are currently evaluating performance of the GPU-compatible binaries, as described in this work.

A major goal of this work is to enable longer AIMD simulations (*i.e.* more time steps) in reasonable wall times. At present, reasonably long (longer than several thousand steps) AIMD simulations are computationally demanding, so practical results are limited and sometimes unattainable. Reducing wall times of AIMD simulations will thus enable higher throughput, larger more realistic models, and the study of additional materials and chemistries that would have been intractable otherwise. As a benchmark for current CPU capabilities, we use a single 128-core CPU node. We then compare the performance of a 4×V100 GPU node to this CPU benchmark. As a figure of merit we use the AIMD production (a number of successful MD steps) produced in 24 hours of wall time.

We note that recently QE was shown to scale well on thousands of CPU cores and thousands of GPU cards [11]. For both architectures, a multi-node setup is required since one node (a motherboard) can hold only a limited number of CPUs and/or GPUs. While system sizes and speedups are intriguing, infrastructure and operational costs of such setups can be significant. We note that effective multi-node parallelization critically depends on fast interconnects such as InfiniBand, HPE Slingshot or Nvidia NVLink switch technologies. However, these are not commonly available on the cloud infrastructure. Similarly, on-premise compute platforms may not have fast interconnection. Thus, in the current study we focus on parallelization strategies and code optimization on a single node of arbitrary size, which is universally available on the cloud and on-premise platforms.

We target a production of 1000 MD steps (or 1 ps with 1 fs time step) as a minimum practically useful throughput in 24 hours running on a single node. Consequently, a single self-consistent field (SCF) calculation of energies and forces should take less than 86 seconds. In this report we study three systems of considerable size and complexity: 1. a free metallic surface; 2. a metal-electrolyte interface; and 3. a metallic grain boundary. The latter two of these systems demonstrate non-trivial behavior (chemical reaction, dissociation, etc.) not attainable with classical MD. To be able to simulate a wide variety of systems (including metallic ones) in a uniform way, we utilize the Born-Oppenheimer MD method as implemented in the QE code.

## 2. General computational details

All AIMD simulations were performed using QE version 7.2 [10] within the NVT ensemble. The time step was set to 1 fs. Ionic temperature was controlled using velocity rescaling to the target temperature at each step. Atomic positions were propagated using the Verlet integration method [12]. The Perdew Burke Ernzerhof (PBE) [13] functional was used to describe the electron exchange and correlation energies within the generalized gradient approximation (GGA) for all systems.

### 2.1 Wall potential constraint implementation

Atomistic modeling of surfaces and interfaces with periodic boundary conditions utilizes slab models whereby two active surfaces or interfaces are present. It is often practical to confine surface processes, such as adsorption, desorption or surface reactions to one of the slab surfaces. To this end, a bias is usually required for molecular dynamics simulations that prevents "runaway" of atoms of interest to another surface. Typically, this is achieved through an increase of the system size in the direction perpendicular to the interface or through introduction of a set of rigid non-reactive atoms (e.g. layer of He atoms [14]). Both methods increase computational cost due to addition of extra space (increase of FFT grid) or additional nuclear cores and electrons. Furthermore, increasing a vacuum gap requires a non-trivial estimation of system size to balance the computation efficiency and accuracy of the calculation.

To enable the simulation of atoms reacting on the selected surface, we have implemented a potential wall constraint in the QE code (available in the QE version 7.2 or later). The potential wall restricts the particle motion through a purely repulsive force in a specified region and acts as an additional boundary condition for atoms. It is defined as:

$$F_{i,z}^{ext} = \begin{cases} A * b * |z_i|^{-b}/z_i & \text{for } z_i < c \\ 0 & \text{otherwise} \end{cases} \quad (1)$$

where $F_{i,z}^{ext}$ is an additional force acting on atom *i* along the *z*-axis, if it is within the cutoff *c* from the origin, *A* is a prefactor, *b* is an exponent, $z_i$ is distance between atom and origin along the z-axis (assumed to be normal to the interface). Cutoff *c*, prefactor *A*, exponent *b* are provided in the input. Due to the polynomial nature of the potential wall implemented, interacting atoms are deflected from the specified boundary. We applied this constraint to the interface systems described in section 3.1.

**2.2 Restartable MD implementation**

Cloud computing enables otherwise prohibitive materials simulations. For example, we previously reported conducting a massive screening of 250,000 charge-conducting organic materials, totaling approximately 3,619,000 DFT single-molecule calculations using Jaguar that took 457,265 CPU hours (~52 years) [15]. As mentioned above, transient nodes present an opportunity to conserve energy compared to the stable nodes. However, their downside being that they can go offline any time (depending on the cloud hardware availability).

Molecular dynamics (MD) has an inherent characteristic of being naturally restartable (continuable) from any point in time given that specific data is saved (in a checkpoint file). In particular, for the Verlet integration method [12], only previous atomic positions are required to successfully continue (restart) an interrupted MD run. Furthermore, our trials show that restarting MD with random wave functions results in a smooth continuity of energies and atomic forces. So, large wave function files are not saved at the checkpoint, thus eliminating the overhead related to management and I/O of these files and minimizing the risk of data loss on occasional node restart of the transient nodes.

We have implemented a job queuing system on top of the most used general queuing systems (SLURM, TORQUE, etc.). Our queuing system can be installed in the cloud by means of a virtual cluster. This setup allows the user to submit a large number of jobs, limited only by the underlying queuing system (e.g. SLURM) scheduling capabilities and cloud resources. Based on this, we have implemented a workflow that runs MD simulation as follows. The driver script runs on a stable node, while MD runs on an entire transient node (Fig. 1). The checkpoint file is continuously synced (using a separate thread, so as not to block the QE calculation) from the transient node to the stable node after each time step. The atomic positions, which are sufficient for restart by the Verlet integrator, are stored in the checkpoint file. The small size of this file makes the time required for I/O and network transfer negligible. This allows us to exclusively run costly DFT jobs on a transient node thus minimizing data loss related to a sporadic node reboot. At the end of the run, all trajectory files are merged into a single trajectory file to ease the analysis.

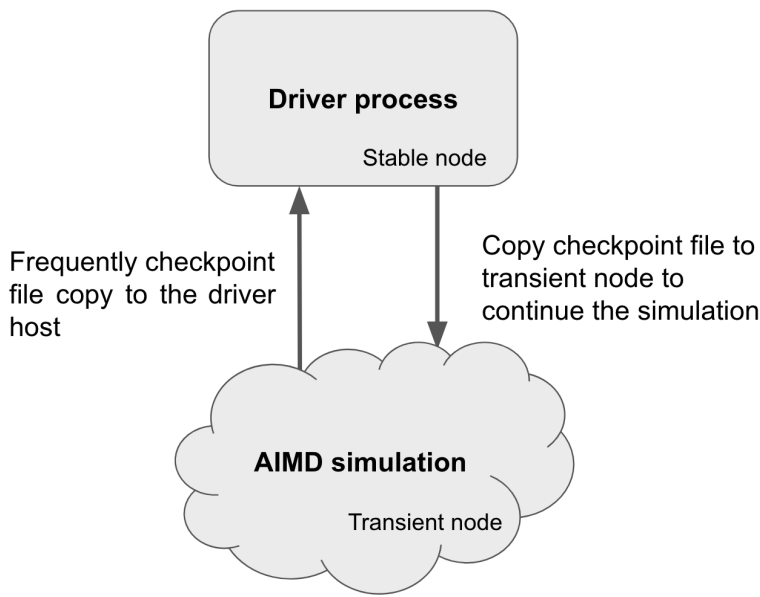

**Fig. 1.** Schematic representation of the restartable AIMD workflow.

**2.3 Massively parallel Bader charge analysis**

Bader charge analysis proved to be instrumental in describing processes where a charge transfer can occur [16]. In the case of an AIMD simulation, it is possible to perform Bader analysis on the trajectory. The scheme of the workflow is shown in Fig. 2. First, equidistant frames are extracted, these are independent and thus can be run in parallel. Our workflow capabilities allow users to set up job graphs with interdependencies. To compute Bader charges, all electron and valence electron charge densities are computed (using the *pp.x* binary of the QE package) and their grid values on an FFT grid are exported, based on the SCF charge density (obtained from the *pw.x* executable). Finally, the Bader analysis program [17] is called to perform charge analysis, which is one of the most commonly accepted measures of charge transfer. Note that if an SCF job fails (due to convergence or other issues), dependent jobs in the graph are not run. For the systems described here, the total wall time of Bader analysis calculations (SCF, valence and total charge densities and Bader) was no longer than 20 minutes, thus we simply implemented an automatic restart from scratch if a server has been preempted.

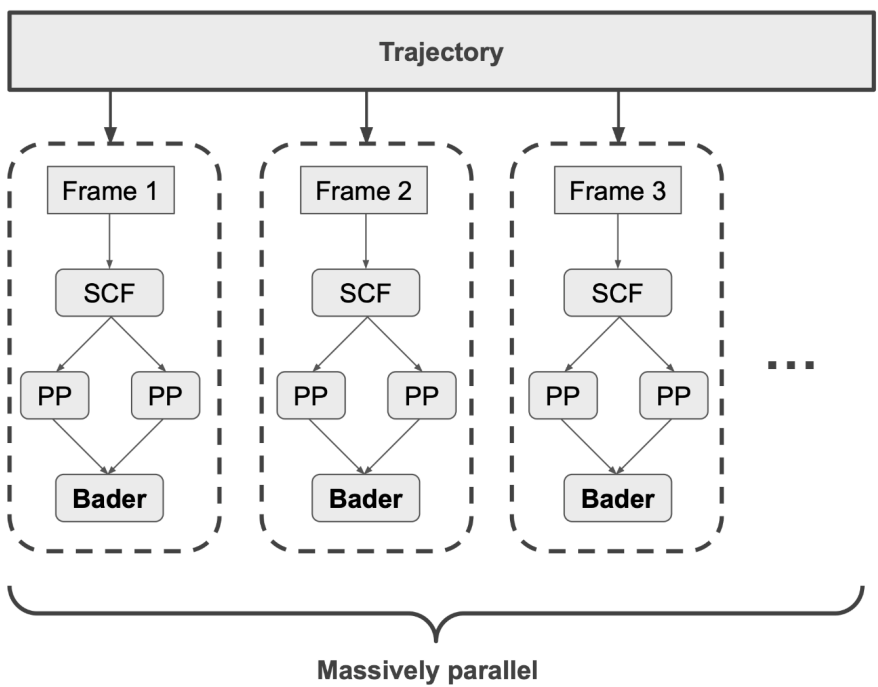

**Fig. 2.** Schematic representation of the workflow to compute Bader charges starting from the AIMD trajectory. Valence and total charge densities are computing using *pp.x* (PP in the figure).

**3. Studied systems and specific calculations setup**

In this section atomic composition of the systems (Table 1) and system specific computation setup details are described.

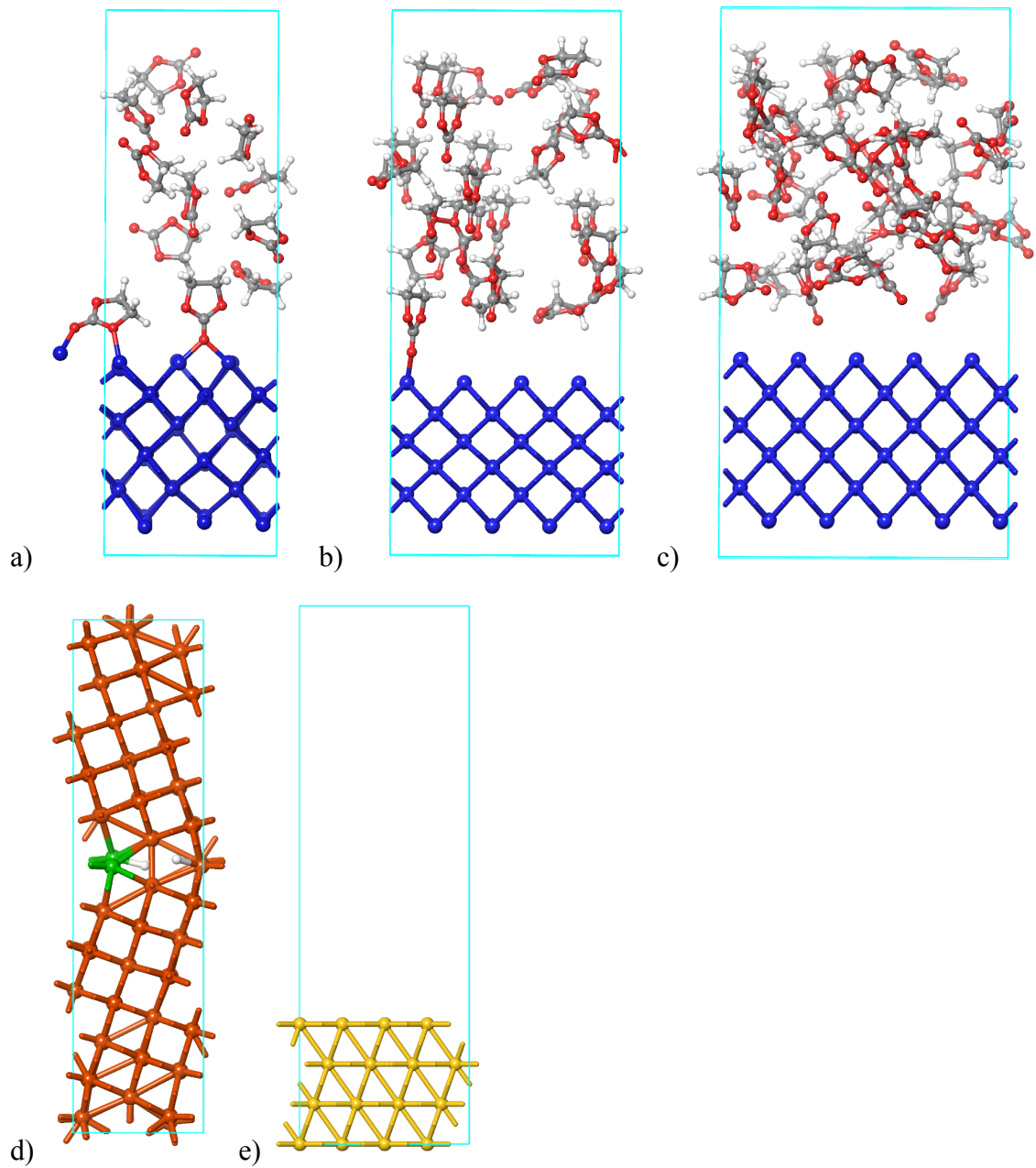

**Fig. 3.** Atomic structures and the unit cells of the systems studied. a) Li(001)-ethylene carbonate interface model (referred as EC-Li(001)); its super cells b) EC-Li(001) [443] and c) EC-Li(001) [553]. Lithium atoms are depicted as blue; carbon - gray, oxygen - red, hydrogen - white; d) Slab model of two hydrogen atoms in the $\Sigma_5$ (100)(0-1-2) tilt grain boundary in BCC iron (H$_2$-FeGB). Fe atoms are depicted as brown and green. H atoms - white; e) Gold (111) surface (Au(111)).

**Table 1.** Composition details of the simulated systems. [443] and [553] refer to the super cell sizes of the Li (001) slab (in x, y and z directions, respectively) in the EC-Li interfacial system. See Fig. 3 and main text.

| System | Composition | Total number of atoms |
|---|---|---|
| EC-Li(001) | $(C_3H_4O_3)_{12}Li_{54}$ | 174 |
| EC-Li(001) [443] | $(C_3H_4O_3)_{23}Li_{96}$ | 326 |
| EC-Li(001) [553] | $(C_3H_4O_3)_{35}Li_{150}$ | 500 |
| $H_2$-FeGB | $Fe_{76}H_2$ | 78 |
| Au(111) | $Au_{112}$ | 112 |

### 3.1 Ethylene carbonate reactive decomposition at Li metal surface

The lithium surface is known to be very reactive. We modeled the reactivity of ethylene carbonate ($C_3H_4O_3$, EC) solvent molecules at the pristine Li(001) surface in the context of Li anode batteries. The AIMD simulations provide important insights into decomposition of EC solvent molecules and identification of key products of the solid-electrolyte interface (SEI) layer formed at the metal anode surface. To benchmark the performance of the GPU implementation of the QE AIMD simulations, several systems of an increasing size (super cells) have been used by varying the interface area and number of the electrolyte molecules (Table 1). The D3 dispersion correction was used for accurate description of the dispersion interactions [18]. Ultrasoft pseudopotentials from the GBRV pseudopotential library were used to describe electron-ion interactions [19]. The energy cut-off of the plane-wave basis expansion was set to 40 Ry (200 Ry for the charge density) with a Gaussian smearing width of 0.01 eV. The gamma-point (with real wavefunctions) was used to sample the Brillouin zone (BZ). The convergence threshold for self consistency was set to $10^{-7}$ Ry. The target ionic temperature was set to 350 K. The wall potential constraint parameters were set to $A$ = 10 Ry/Bohr, $b$ = 2, $c$ = 2.835 Bohr for all the interfaces. The parameters depend on the system, temperature, and time step used. For instance, with increasing temperature and/or time step, the cutoff value *c* should be increased to ensure that atoms do not cross the barrier in one time step. The current barrier parameters were determined empirically and parameterized for a temperature of 300 K and 1 fs time step.

**3.2 Hydrogen at the ferromagnetic iron grain boundary**

Hydrogen aggregation in metals even in small concentrations causes embrittlement and fracture failure. Many proposed mechanisms for hydrogen related embrittlement are believed to be caused by H aggregation and diffusion along the grain boundaries. Such mechanisms are typically addressed using classical MD simulations [20] or using static DFT approaches [21]. The MD simulations typically address time evolution and thermodynamics of relatively large systems, although the quality of the classical force fields and their influence on the results is still a matter of debate. At the same time DFT calculations typically represent particle interactions more accurately, but they are computationally expensive and usually address energies and activation energies of individual atoms, thus neglecting correlations in atomic dynamics. DFT applications for light atom dynamics in ferromagnetic systems are still rare. Here we demonstrate that AIMD can be readily used for such systems, thus capturing both accurate energetics and collective dynamic behavior.

The BCC iron $\Sigma_5$ (100)(0-1-2) tilt grain boundary structure was generated using the Pymatgen python library (Fig. 1) [22]. The structure files are also available from the Grain Boundary Database (GBDB) [23]. A 2×1×1 supercell of the initial unit cell was used in the calculations that contains 76 iron atoms. In the initial configuration, an $H_2$ molecule at equilibrium distance was placed in the largest void in the GB region, so as to maximize H-Fe distances. Spin-polarized calculations were used with initial magnetisation set to ferromagnetic ordering. The energy cutoff of the plane-wave basis expansion was set to 30 Ry (240 Ry for the charge density) with a Marzari-Vanderbilt-DeVita-Payne (MV) cold smearing [24] width of 0.02 eV. The convergence threshold for self consistency was set to $10^{-5}$ Ry. The BZ integration was performed using the Monkhorst-Pack method with 4×4×1 k-point grid, resulting in 8 k-points. The target ionic temperature was set to 300 K. These somewhat loose parameters were found to be sufficient to obtain accurate atomic forces and stable molecular dynamics runs.

### 3.3 Gold surface

The gold surface (Au(111)) slab is one of the standard QE benchmarks. It is composed of 112 gold atoms (4 atomic layers) with a vacuum gap of 25 Å along the z-axis. Starting computation parameters were identical to those provided in the QE benchmarks repository [25]. Ultrasoft pseudopotentials from the benchmark repository were used to describe electron-ion interactions. The energy cutoff of the plane-wave basis expansion was set to 25 Ry (200 Ry for the charge density) with a MV cold smearing width of 0.05 eV. The convergence threshold for self consistency was set to $10^{-5}$ Ry. The BZ integration was performed using Monkhorst-Pack method with a 2×2×1 k-point grid, resulting in 2 k-points. The target ionic temperature was set to 350 K.

## 4. Results

### 4.1 Performance discussion

We ran AIMD simulations on both GPU and CPU nodes (see Table 2, Fig. 4). The GPU node had 60 GB RAM, a total number of 16 CPU cores (Intel(R) Xeon(R) CPU @ 2.30GHz), and 4 Nvidia(R) v100 GPUs with 16 GB GPU RAM each. GPU calculations were run on the entire node; for each GPU (MPI process), 4 CPUs using OpenMP were employed. The CPU node had 516 GB RAM and a total number of 128 CPU cores (Intel(R) Xeon(R) CPU @ 2.60GHz). We note that in both CPU and GPU cases, we used the entire node, so that no other processes interfere with the calculations. The productivity of the AIMD runs was measured as a number of MD steps calculated in 24 hours.

For EC-Li(001) systems Gamma-point only calculations were run using automatic MPI parallelization (available in QE version 7.1 or later). For the $H_2$-FeGB system (with BZ sampled using 8 k-points), on the GPU node, calculations were further parallelized over k-points (*-npools* = 4, i.e 2 k-points per GPU). For the Au(111) surface (with BZ sampled using 2 k-points), one k-point on two GPUs (*-npools* = 2).

The Au(111) surface was previously benchmarked using machines with different GPU architectures [26]. Our timing for the single point calculation on the GPU node agrees with these benchmarks. We have also optimized the *mixing_beta* setting and found that by reducing its value from 0.7 (QE default) to 0.1, the average number of iterations needed for SCF convergence was reduced from 13 to 9. This resulted in a ~30% speedup of MD simulations (see Table 2).

It is instrumental to compare performance between GPU and CPU nodes. For the EC-Li(001) system, the simulation on the GPU node was twice more productive compared to the CPU node: 5486 *vs* 2497 MD steps per 24 hours. Even better GPU performance improvement is seen for the EC-Li(001) [443] system: 1753 *vs* 564 MD steps. We note however that GPU nodes are limited by the GPU RAM of a single card. This is a hard restriction, which is much less critical in CPU nodes where more RAM can be added to the motherboard. We found that calculations of the [553] extension of the EC-Li(001) interface (500 atoms) run out of GPU RAM and fail on the GPU node, whereas it runs successfully on the CPU node. For the Au(111) and $H_2$-FeGB systems, the GPU outperforms the CPU more than three times. We note that in these two systems, heavier atoms are present with more valence electrons compared to the EC-Li systems. Thus, if GPU nodes are available and the system can fit in the GPU RAM, GPU nodes offer a significant productivity improvement over CPU nodes.

Using the restart mechanism described in section 2.2, we performed three consecutive runs (each limited to 8 hours of wall time) on the GPU node, emulating the transient node going offline twice, totaling 24 hours of simulation wall time. The restart functionality enables users to run extended AIMD simulations on transient nodes without losing any data. For all the systems, three consecutive runs resulted in a similar number of the time steps compared to the uninterrupted runs, within a 10% statistical error (see Table 2). The total time, which we define as wall time + queuing time, can increase greatly on the transient servers compared to the stable one; on a stable server, the calculation is queued only once, whereas the number of times a transient server will go offline and a new instance will be requested is unpredictable.

**Table 2.** AIMD simulation runs in steps per day (a day is defined as 24 hours of simulation wall time, not total time). Thus, displayed run times reflect times taken by the MD production only, no post-processing or queue time.

| System | 1 run on CPU node | 1 run on GPU node | 3 consecutive runs (8 hrs each) on GPU node |
|---|---|---|---|
| EC-Li(001) | 2497 | 5486 | 5814 |
| EC-Li(001) [443] | 564 | 1753 | 1910 |
| EC-Li(001) [553] | 215 | Out of GPU RAM | Out of GPU RAM |
| $H_2$-FeGB | 127 | 796 | 801 |
| Au(111) (mixing_beta = 0.7) | 321 | 1649 | 1637 |
| Au(111) (mixing_beta = 0.1) | 379 | 2175 | 2204 |

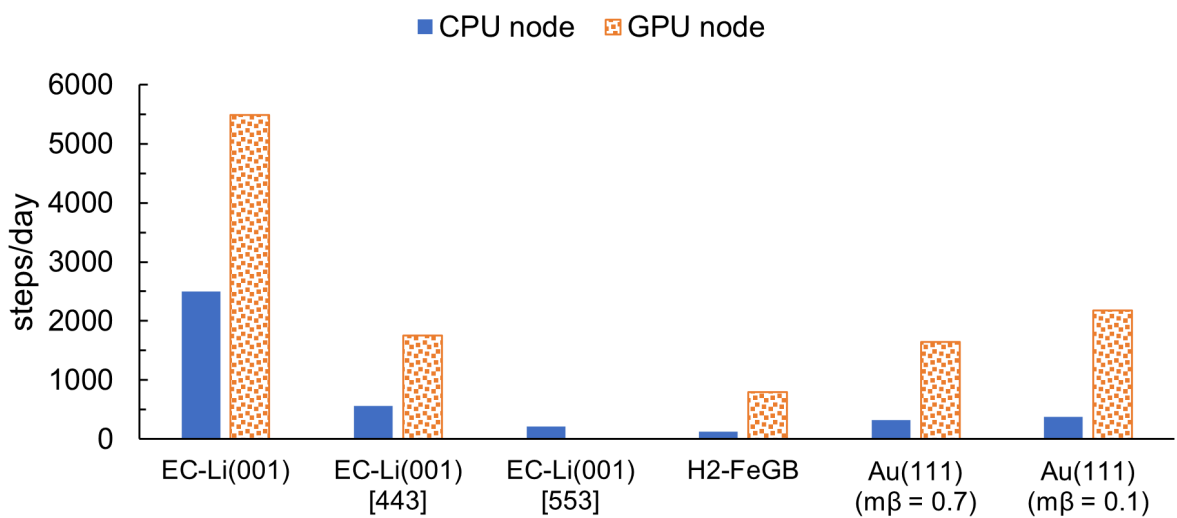

**Fig. 4.** CPU and GPU performance in steps per day for the systems under consideration (data from columns 1 and 2 of Table 2).

## 4.2 Dynamics discussion

In this section we discuss scientific results of our Li/EC interface and iron grain boundary AIMD simulations. In both cases, we observe interesting reactive phenoma, demonstrating that the simulation throughput achieved by our GPU calculations is sufficient.

### 4.2.1 Reactivity at the Li/EC interface

Trajectory analysis of the EC-Li(001) system shows reactive dynamics, indicating that the simulation time is sufficient to capture phenomena of interest. Specifically, the first EC molecule decomposes at 0.2 ps, leading to the formation of an ethylene ($C_2H_4$) molecule and a carbonate ion ($CO_3^{2-}$) adsorbed at the Li surface as illustrated in Fig. 5 (a). Similarly, at 0.6 ps another EC molecule undergoes a decomposition reaction forming a bound CO molecule as illustrated in Fig. 5 (b). The ring-opened EC byproduct adsorbs at the Li surface and remains stable for the remaining duration of the AIMD simulation. Both the reaction mechanisms observed in our AIMD simulations are in agreement with results reported in the literature [27, 28, 29].

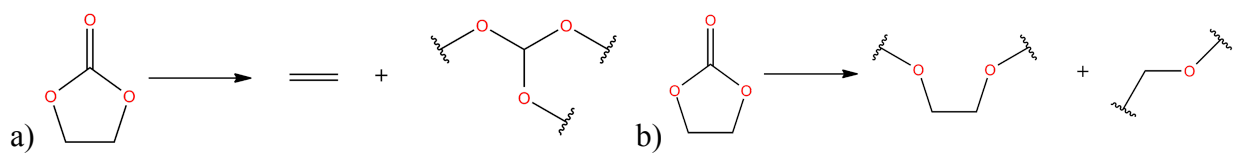

**Fig. 5.** Two different reactions observed during the EC-Li(001) AIMD simulation. The wiggle bond indicates binding to the Li surface.

Bader charge analysis is instructive in quantification of charge transfer in the reactions occurring at the interface. The Bader charge was computed for each atom of the Li surface slab and EC solvent molecules. The time evolution of the change in the per-atom Bader charge transferred from the Li (001) surface slab to the solvent molecules is shown in Fig. 6. Here, the total per atom Bader charge of the Li (001) slab at 0 ps is taken as a reference. As seen from the plot, the charge transfer from the Li surface slab saturates at a value of ~0.08 $e$/atom within 1 ps of the AIMD simulation time indicating rapid reactivity and decomposition of the two EC solvent molecules. The net charge of the Li slab fluctuates around a constant value for the remaining simulation duration indicating no further reaction of the EC molecules with the Li surface.

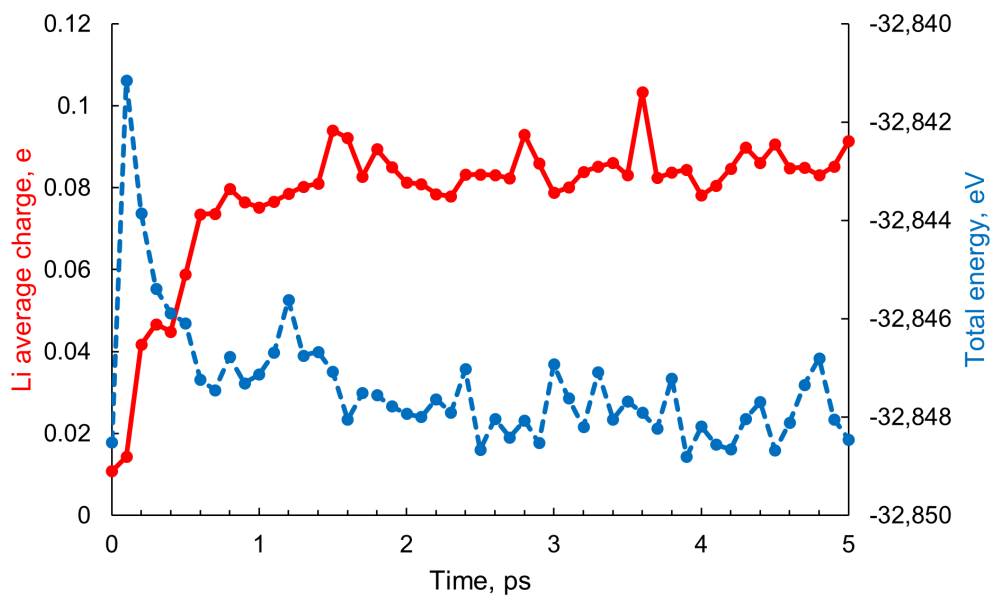

**Fig. 6.** Time dependence of the averaged charge of the Li ions (left y-axis, red solid line) and total energy (right y-axis, blue dashed line).

**4.2.2 Hydrogen dissociation at the iron grain boundary**

AIMD for hydrogen in the iron grain boundary presents a wealth of information going beyond the scope of this paper. Let us summarize here just a few aspects that follow from the trajectory analysis and are not typically attainable in classical MD or static DFT calculations:

1. The grain boundary Fe atoms (Fig. 3(d) highlighted green) exhibit spin localisation whereby the atomic spin is 2.2 Bohr compared to the average Fe spin of 1.8 Bohr.
2. As expected two H atoms do not form a molecule, but their motion is highly correlated resulting in a loose $H_2$ agglomerate with a typical R(H-H) distance varying between 2.4 and 2.9 Å (Fig. 7)
3. Hydrogen is confined to the grain boundary and exhibits rapid diffusion.
4. The apparent H activation energy (not discussed in the detail) is significantly lower than

that predicted from the static DFT for a single H atom [21].

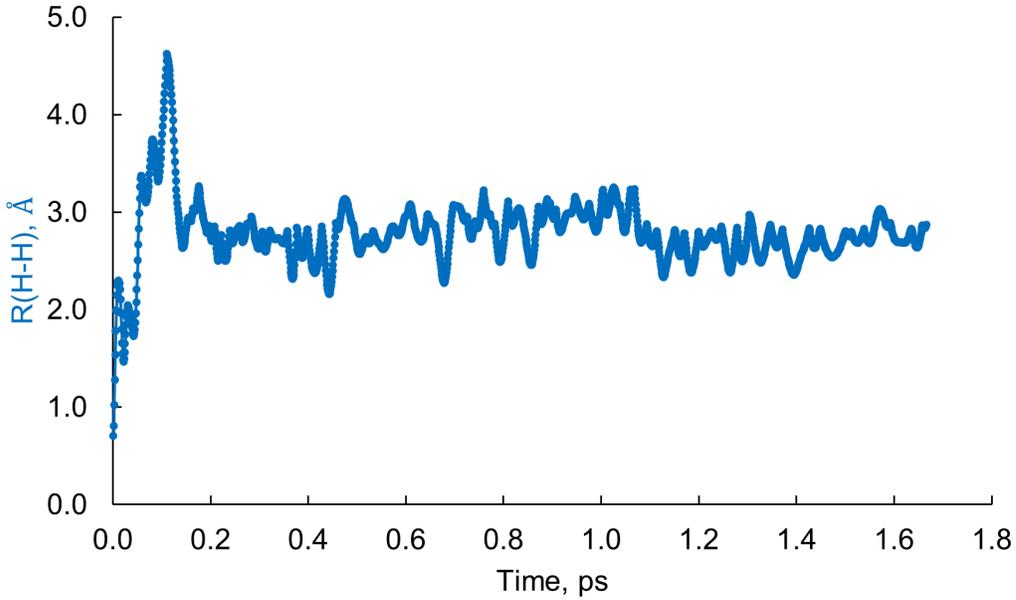

**Fig. 7.** Hydrogen-hydrogen distance as a function of time while diffusing along the grain boundary.

## 5. Conclusions

*Ab initio* molecular dynamics unquestionably leads to a broader class of problems addressable by atomistic modeling. However, such calculations are computationally expensive, which often restricts such modeling. In this paper we argue that in order to be instructive for a specific system, AIMD must be capable of generating at least 1000 MD steps in 24 hours on a moderate computational resource. With this goal in mind we described the implementation of a restartable *ab initio* Born-Oppenheimer MD calculations using Quantum ESPRESSO as the periodic DFT backend. We showed that calculations executed on transient GPU or CPU nodes lead to required performance.

Towards more effective interface modeling, we have implemented a wall potential constraint restricting atomic dynamics to one of the slab plains. This allows modeling systems with smaller vacuum gaps thus reducing computational expense and increasing productivity.

We performed AIMD of three diverse systems, two of which show non-trivial reactive dynamics. We compared timings for a single node CPU and GPU architectures in a cloud environment. As previously reported by other groups, GPU implementation provides a significant computational advantage. The described approach allows one to perform routinely realistic calculations (around 1 ps or longer) given time (24 hours) and infrastructure constraints (single CPU or GPU node). The discussed implementation is available via the Schrödinger Materials Science Suite [9], which provides a graphical environment for atomistic modeling, including advanced structure generation, setup, visualization and analysis, and automated workflows.

**Credit authorship contribution statement**

A. Fonari: Writing original Draft, Methodology, Software. G. Agarwal: Data curation, Writing original Draft, Validation. S. C. Tiwari: Methodology, Software. C. N. Brock: Methodology, Visualization. J. Gavartin: Validation, writing - Original Draft. M. D. Halls: Writing - Reviewing and Editing,

**Declaration of competing interest**

The authors declare that they have no known competing financial interests or personal relationships that could have appeared to influence the work reported in this paper.

**Acknowledgements**

Authors are grateful to Pietro Davide Delugas and Paolo Giannozzi for review of the wall potential code and general expert advice on Quantum ESPRESSO code. Louis Stuber and Filippo Spiga of Nvidia(R) for expert advice and help in compilation and optimisation of the Quantum ESPRESSO GPU enabled binaries. Funding from Gates Ventures is gratefully acknowledged.

**Data availability**

The data required to reproduce these findings are available from the authors on request.